\documentclass{article}
\usepackage{spconf,amsmath,graphicx}
\usepackage{amssymb}
\usepackage{mwe} 
\usepackage[colorlinks,linkcolor=blue]{hyperref}

\title{A Unified Conditional Disentanglement Framework for Multimodal Brain MR Image Translation}
\name{Xiaofeng Liu, Fangxu Xing, Georges El Fakhri, Jonghye Woo}
\address{Dept. of Radiology, Massachusetts General Hospital and Harvard Medical School, Boston, MA, USA
}
\begin{document}
%
\maketitle

\begin{abstract}\vspace{-5pt}
Multimodal MRI provides complementary and clinically relevant information to probe tissue condition and to characterize various diseases. However, it is often difficult to acquire sufficiently many modalities from the same subject due to limitations in study plans, while quantitative analysis is still demanded. In this work, we propose a unified conditional disentanglement framework to synthesize any arbitrary modality from an input modality. Our framework hinges on a cycle-constrained conditional adversarial training approach, where it can extract a modality-invariant anatomical feature with a modality-agnostic encoder and generate a target modality with a conditioned decoder. We validate our framework on four MRI modalities, including T1-weighted, T1 contrast enhanced, T2-weighted, and FLAIR MRI, from the BraTS'18 database, showing superior performance on synthesis quality over the comparison methods. In addition, we report results from experiments on a tumor segmentation task carried out with synthesized data.\footnote{Published in ISBI 2021 for Oral presentation.}
\end{abstract}\vspace{-5pt}
\begin{keywords}
Image synthesis, Generative Adversarial Networks, Deep learning, Brain tumor
\end{keywords}\vspace{-5pt}

\section{Introduction}\vspace{-5pt}
\label{sec:intro}
Multimodal magnetic resonance (MR) images are often required to provide complementary information for clinical diagnosis and scientific studies \cite{armanious2020medgan,liu2020symmetric}.~For example, multimodal MR imaging (MRI) with T1-weighted, T1ce (contrast enhanced), T2-weighted, and FLAIR (FLuid-Attenuated Inversion Recovery) MRI can offer greater sensitivity to tumor heterogeneity and growth pattern than single modality, T1ce MRI, thereby benefiting diagnosis, staging, and monitoring of brain metastasis~\cite{chang2018subject}. However, in practice, it is often difficult to acquire sufficiently many modalities due to limitations in study plans, and some modalities could be missing due to imaging artifacts~\cite{dar2019image,yu2019ea}.

In recent years, cross-modality synthesis of brain MR images using generative adversarial networks (GANs) has gained its popularity \cite{armanious2020medgan}. For example, Yu et al. \cite{yu2019ea} adopted a pair-wise image-to-image network via Pix2Pix \cite{isola2017image,liu2020symmetric} for transferring T1-weighted to either T2-weighted or FLAIR MRI. Also, a cycle-reconstruction approach via CycleGAN for unpaired image translation \cite{zhu2017unpaired} was introduced in \cite{dar2019image,qu2020multimodal} to stabilize the training. These methods \cite{armanious2020medgan,yu2019ea,dar2019image,qu2020multimodal,liu2020symmetric} aimed at modeling the mapping between two specific modalities, requiring two inverse autoencoders to achieve the cycle-reconstruction \cite{zhu2017unpaired}.

However, the aforementioned approaches have a limitation in that they dealt with the cross-modality synthesis problem which cannot be easily scalable to multiple modalities (i.e., more than two modalities).~In other words, in order to learn all mappings among $M$ modalities, $M(M-1)$ different generators have to be trained and deployed (e.g., 12 possible cross-modality networks for T1-weighted, T1ce, T2-weighted, and FLAIR MRI) \cite{menze2014multimodal}. Moreover, each translator cannot fully use the entire training data, but can only learn from a subset of data (two out of $M$ modalities). Failure to fully use the whole training data is likely to limit the quality of generated images.~To address this, recently, Xin et al.~\cite{xin2020multi} proposed to construct a 1-to-3 network to translate T2-weighted to T1-weighted/T1ce/FLAIR MRI based on Pix2Pix \cite{isola2017image}. The improvement over Pix2Pix was achieved by utilizing 3$\times$ training pairs for one translator \cite{xin2020multi}. Besides, the closely related multiple tasks mutually reinforced each other \cite{choi2018stargan}. Yet, with the 1-to-3 network, the number of models to be trained was still limited to the number of modalities.

In this paper, we propose to achieve all of the pair-wise translation using a single set of conditional autoencoder and discriminator. Our framework is scalable to many modalities and can effectively use all of possible paired cross-modality training data. Several unpaired multi-domain synthesis methods are inherently multimodal translation, while they usually require multiple domain-specific autoencoders \cite{huang2018multimodal} and discriminators \cite{yu2018singlegan}, and do not consider pair-wise training data \cite{huang2018multimodal,yu2018singlegan,yuan2019unified,choi2018stargan}.~Without the pair-wise supervision, the largely unconstrained image generation tends to alter important characteristics of an input modality for generating diverse outputs.~Unlike image-to-image translation in computer vision, in medical domain, it is of paramount importance not to introduce unnecessary changes and artifacts to carry out quantitative analyses \cite{siddiquee2019learning}. 

In addition, these methods ignore the inherent connection between different MR modalities \cite{dewey2020disentangled,dewey2020disentangled}. Since multiple MR modalities are acquired with different scan parameters for the same subject, there should be a shared modality-invariant anatomical feature space \cite{dewey2020disentangled}. Accordingly, we propose to configure a pair-wise disentanglement approach \cite{liu2019feature,mathieu2016disentangling} to extract an anatomical feature with a modality-agnostic encoder, and then inject a modality-specific appearance with a conditional decoder.

Specifically, our unified multimodal translation framework hinges on the autoencoder, where the decoder is conditioned on the target modality to utilize the paired training data. After the feedforward processing of the autoencoder conditioned on any modality label, the same autoencoder is called again, which conditioned on the modality label of the original input for the cycle-reconstruction. The anatomical information disentanglement is simply enforced by the similarity of the output feature map of the encoder \cite{mathieu2016disentangling,liu2019feature}.


We empirically validate its effectiveness on the BraTS'18 database, showing superior performance over the comparison methods.

\vspace{-15pt}

\section{Methodology} \vspace{-5pt}

\begin{figure}[t]
\begin{center}
\includegraphics[width=1.0\linewidth]{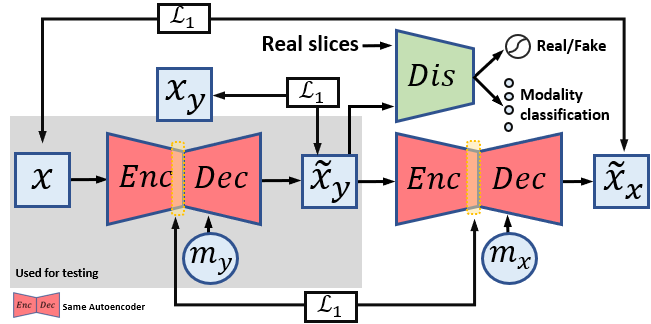}\vspace{-15pt}
\end{center} 
\caption{Illustration of the proposed unified conditional adversarial framework for multimodal co-registered brain MR image translation. Note that only one $Enc$-$Dec$ set is recalled twice, and only the gray masked subnets are used for testing.} \vspace{-10pt}
\label{fig:illus}
\end{figure}

Given a set of co-registered $M$ MR modalities, a sample $x$ with modality $m_x$ has $M-1$ pixel-wise aligned samples with the other modalities. The target modality of image synthesis and the corresponding ground truth sample are denoted as $m_y$ and $x_y$, respectively. Given the pair of input sample and target modality $\{x,m_y\}$, we propose to learn a parameterized mapping $f:\{x,m_y\} \rightarrow \tilde{x}_{y}$ from $\{x,m_y\}$ to the generated corresponding sample with modality $m_y$ to closely resemble $x_y$. $m_y$ denotes a four-dimensional one-hot vector to represent the four MR modalities available in the BraTS'18 database. Of note, $m_y=m_x$ indicates the self-reconstruction.~The proposed framework is shown in Fig.~\ref{fig:illus}.
\vspace{-5pt}
\subsection{Disentangled United Multimodal Translation}\vspace{-5pt}

A straightforward baseline network structure for paired two-modality translation is an autoencoder, which is constructed with an encoder $Enc$ and a decoder $Dec$. Briefly, it first maps $x$ to a latent feature $z$ via the $Enc$, and then decode $z$ to reconstruct the target image via the $Dec$. The target ground truth image serves as a strong supervision signal, while the unpaired translation cannot benefit from such regularization \cite{zhu2017unpaired,liu2017unsupervised,huang2018multimodal,choi2018stargan,yu2018singlegan}.

However, the autoencoder has a limitation in that the generated images are likely to be blurry \cite{liu2020symmetric}, which is partly caused by the element-wise criteria such as the $\mathcal{L}_1$ or $\mathcal{L}_2$ loss \cite{larsen2015autoencoding}. Although recent studies \cite{kingma2016improved} have substantially improved the predicted log-likelihood in the autoencoder, the image generation quality of the autoencoder is still inferior to GANs. In addition, enforcing pixel-wise similarity is likely to distract the autoencoder from understanding the underlying anatomical structure, when inputting slightly misregistered data.

In order to enforce high-level semantic similarity and improve quality of generated textures, recent cross-modality translation models \cite{dar2019image,yu2019ea,xin2020multi} adopted an additional adversarial loss $\mathcal{L}_{adv}$ with a discriminator $Dis$ following Pix2Pix \cite{isola2017image,liu2020symmetric}, where the training objectives consist of both the $\mathcal{L}_1$ loss and adversarial GAN loss:\vspace{-5pt}\begin{align}
    ^{\text{~~~~~min}}_{Enc,Dec} ~^{\text{max}}_{Dis}~~&\mathcal{L}_1(\tilde{x}_{y},x_y)=|\tilde{x}_{y}-x_y|,\\
    ^{\text{~~~~~min}}_{Enc,Dec} ~^{\text{max}}_{Dis}~~&\mathcal{L}_{adv}= \mathbb{E}\text{log}(Dis(\tilde{x}_{y})) + \mathbb{E}\text{log}(1-Dis(\tilde{x}_{y})).
\end{align}

\vspace{-10pt}To extend the Pix2Pix basenet to multimodal translation, we adopt the conditional decoder structure that takes both the feature map extracted by the encoder $Enc(x)$ and the target modality code $m_y$ as input. The target modality code is spatially replicated and concatenated with the input image. Of note, the unpaired multi-domain synthesis network takes the modality code as input to the encoder \cite{huang2018multimodal,yu2018singlegan,yuan2019unified,choi2018stargan}; therefore it cannot achieve the disentanglement of modality-agnostic anatomic and modality-specific factors \cite{dewey2020disentangled,liu2019feature,mathieu2016disentangling}. Then, a single autoencoder model can be switched to all possible pair-wise cross-modality translations. Therefore, $\tilde{x}_{y}$ in Eqs. (1-2) can be $Dec(Enc(x),m_y)$.

Instead of configuring $M$ $Dis$ for all modalities, we introduce an auxiliary modality classifier $Dis_{mc}$ \cite{odena2017conditional} on top of $Dis$ that allows a single $Dis$ to control multiple modalities. The to be minimized modality classification loss can be formulated as:\vspace{-5pt}\begin{align}
\mathcal{L}_{mc}=^\mathbb{E}_{\sim\tilde{x}_{y}} [-\text{log} Dis_{mc} (\tilde{x}_{y},m_y)]+^\mathbb{E}_{\sim x} [-\text{log} Dis_{mc} (x,m_x)].
\end{align}\vspace{-10pt}

\begin{figure}[t]
\begin{center}
\includegraphics[width=1.0\linewidth]{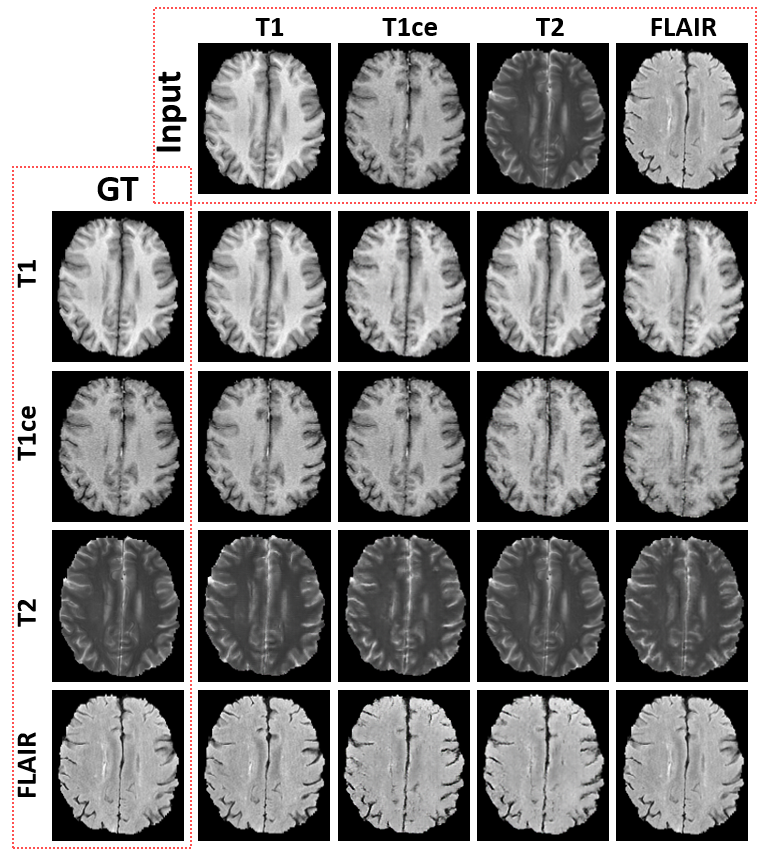}\vspace{-20pt}
\end{center} 
\caption{Illustration of the results of our proposed framework. We use the first row as input, and configure four target modalities. The diagonal results are obtained using self-translation (i.e., $m_y=m_x$).}\vspace{-10pt}
\label{fig:result1}
\end{figure} 

The objective of conditional GAN with multi-task discriminator can induce an output distribution of over $(\tilde{x}_{y}|m_y)$ that matches the empirical distribution of real images with modality $m_y$, i.e., $x_y$. However, the mapping between two distributions can be largely unconstrained and have many possible translations $f$ to induce the same distribution over $f(\{x,m_y\})$ \cite{zhu2017unpaired}. Therefore, the learned $f$ cannot guarantee that the individual inputs $\{x,m_y\}$ and outputs $\tilde{x}_{y}$ are paired up as expected. To mitigate this, CycleGAN \cite{zhu2017unpaired} is proposed to introduce an additional cycle-reconstruction constraint for unpaired two-domain image translation.~Specifically, the generated output is mapped back to the original image with an inverse translator, and the $\mathcal{L}_1$ loss is explicitly used as a loss function to measure the similarity between the mapped back image and the original input.~In this way, the shape structure can be better maintained. Note that both Pix2Pix \cite{isola2017image} and CycleGAN \cite{zhu2017unpaired} are used as the two-domain translators, and there are two inverse autoencoders to achieve the cycle reconstruction \cite{zhu2017unpaired}.  

In addition, rather than configuring two autoencoders with inverse direction \cite{zhu2017unpaired}, we can simply recall the same autoencoder twice with a different conditional modality code in a feedforward processing. Specifically, the second time translation always uses the modality of original input sample $m_x$ to achieve the reconstruction of $x$, given by\vspace{-5pt}\begin{align}
    ^{\text{~~~~~min}}_{Enc,Dec}~~&\mathcal{L}_1(\tilde{x}_{x},x_x)=|\tilde{x}_{x}-x|,
\end{align} where $\tilde{x}_{x}=Dec(Enc(\tilde{x}_{y}),m_x)$ is expected to reconstruct $x$.


To achieve the disentanglement of anatomical information and modality-specific factors without the anatomical label, we propose to enforce the similarity between $Enc(x_x)$ and $Enc(\tilde{x}_{x})$ which are the two encoder outputs in a cycle, given by \vspace{-5pt}\begin{align}
    ^{\text{min}}_{Enc}~~&\mathcal{L}_1^{disen}=|Enc(\tilde{x}_{y})-Enc(x)|,
\end{align} which explicitly requires that the paired co-registered two-modality images have the same feature map. Their shared factors can be the anatomical information \cite{dewey2020disentangled}, and have the difference between modality-specific imaging parameters \cite{dewey2020disentangled}. The feature vector is also required to be combined with the target modality label to reconstruct the target image, which encourages them to be independent and complementary to each other \cite{liu2019feature}. Therefore, the latent feature space can be anatomically related and modality-invariant (i.e., disentangled with a modality factor) \cite{mathieu2016disentangling,liu2019feature}. In addition, the feature-level similarity is also related to the perception loss \cite{liu2019hard}, which enhances the textures.  
\vspace{-5pt}
\subsection{Training Strategy}\vspace{-5pt}
For simpler implementation, we reformulate the min-max terms to minimization only in a consistent manner. The objective of our conditional autoencoder and the adversarial cycle-reconstruction streams can be formulated as:\vspace{-5pt}\begin{align}
    &^{\text{~~~~~~~min}}_{Enc,Dec} \mathcal{L}_1(\tilde{x}_{y},x_y)+\alpha \mathcal{L}_1(\tilde{x}_{x},x_x)+\beta \mathcal{L}_{adv}+\lambda_1\mathcal{L}_{mc},\\
    &~~~~~^{\text{min}}_{Dis} -\mathcal{L}_{adv}+\lambda_2 \mathcal{L}_{mc},\vspace{-10pt}
\end{align} where $\alpha$, $\beta$, $\lambda_1$, and $\lambda_2$ are the weighting hyperparameters. $Enc$ and $Dec$ minimize $\mathcal{L}_{adv}$, while $Dis$ minimize $-\mathcal{L}_{adv}$ to play a round-based adversarial game to improve each other to find a saddle point. In practice, we sample the same number of $x$ from each modality in training \cite{isola2017image}. 
\vspace{-5pt}
\subsection{Testing Translation}\vspace{-5pt}
After training, we can obtain the translation functions by assembling a subset of the subnetworks, i.e., $Enc$ and $Dec$. Note that our translation can be agnostic to an input modality. Therefore, with an input sample $x$ and a target modality $m_y$, we can generate its corresponding $\tilde{x}_{y}=Dec(Enc(x),m_y)$. With generated images with a target modality, we concatenate them for further tumor segmentation~\cite{xin2020multi,qu2020multimodal}.

\begin{figure}[t]
\begin{center}
\includegraphics[width=1.04\linewidth]{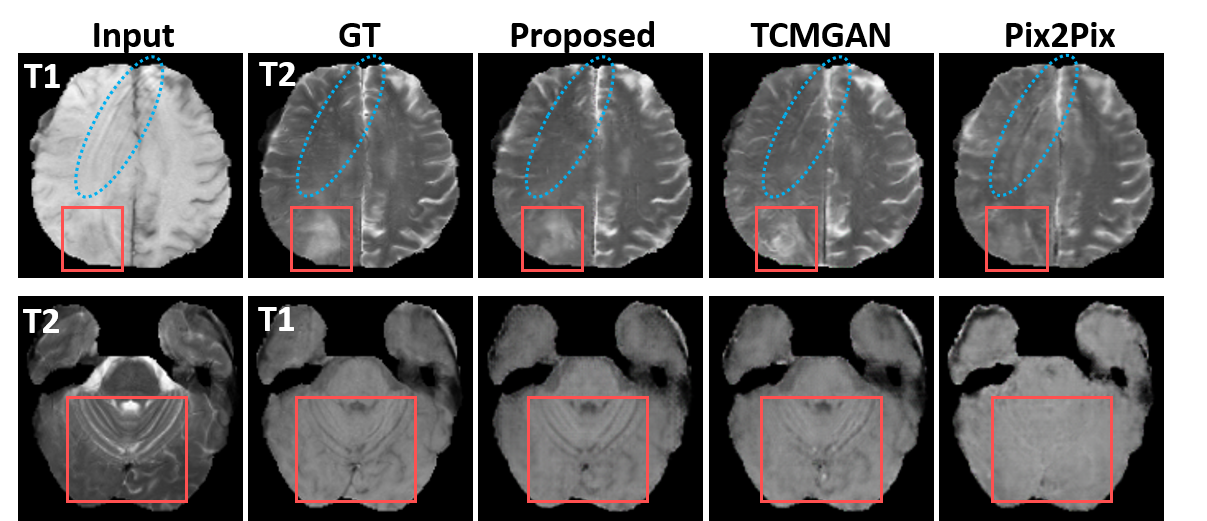}\vspace{-20pt}
\end{center} 
\caption{Comparison of different methods for the T1-weighted and T2-weighted MR translation.~GT indicates the ground truth $x_y$.}\vspace{-15pt}
\label{fig:result2}
\end{figure} 

\section{RESULTS AND DISCUSSION}\vspace{-5pt}
 
We evaluated our framework on the BraST'18 multimodal brain tumor database \cite{menze2014multimodal}, which contains a total of 285 subjects with four MRI modalities: T1-weighted, T1ce, T2-weighted, and FLAIR MRI, with the size of 240$\times$240$\times$155. The intensity of slices was linearly scaled to [$-1,1$] as in \cite{xin2020multi,yu2019ea}, which was then processed by 2D networks. The axial slices with less than 2,000 pixels in the brain area were filtered out as in~\cite{xin2020multi}.

For a fair comparison, we followed \cite{xin2020multi} to use 100 subjects for the training translator, 85 subjects for testing, and 100 subjects for a training segmentor. We adopted the same backbone for $Enc$, $Dec$, and $Dis$ for all comparisons \cite{xin2020multi}. In addition, the reimplemented Pix2Pix \cite{isola2017image} was used as the two-modality transfer baseline model. In order to align the absolute value of each loss, we set weights $\alpha=1$, $\beta=0.5$, $\lambda_1=1$, and $\lambda_2=1$. We used Adam optimizer with a batch-size of 64 for 100 epochs training. The learning rate was set at $lr_{Enc,Dec}=1\mathrm{e}{-3}$ and $lr_{Dis}=1\mathrm{e}{-4}$ and the momentum was set at 0.5. Our framework was implemented using PyTorch.~The training on an NVIDIA V100 GPU took about 8 hours.~In practice, translating one test image with our unified $Enc$ and $Dec$ only took about 0.1 seconds. 

\vspace{-5pt}
\subsection{Qualitative Evaluations}\vspace{-5pt}

In Fig.~\ref{fig:result1}, we illustrated the multimodal generation results of 12 cross-modality translation tasks and 4 self-reconstruction tasks. The proposed framework successfully synthesized any modality by simply configuring the target modality, which is consistent with the target ground truth MR images. We note the self-supervision was not used for the subsequent segmentation, but was used for checking the image generation quality in our implementation.

The qualitative comparisons with the 1-to-1 translator Pix2Pix \cite{isola2017image} and 1-to-3 translator \cite{xin2020multi} are shown in Fig.~\ref{fig:result2} along with our proposed framework. The proposed framework was able to generate visually pleasing results with better shape and structure consistency when visually assessed. From the red box in the first row, we can see that the tumor area was better maintained with the help of the cycle-constraint compared with \cite{xin2020multi} which uses the additional tumor-consistent loss.~Also, the artifact shown in the T1-weighted MR image (i.e., stripes indicated by the blue circle) yielded similar stripes as shown in the T2-weighted MR image with TCMGAN and Pix2Pix. Our disentangled encoder was able to eliminate the artifact and enforce the latent representation following the distribution of normal MR images.

\begin{table}[t]
\centering
\caption{Numerical comparisons of four methods in testing}
\resizebox{1\linewidth}{!}{
\begin{tabular}{c|c|c|c|cccc|ccc}
\hline
Methods& L1~$\downarrow$&SSIM~$\uparrow$ &	PSNR~$\uparrow$ 	&  IS~$\uparrow$\\\hline\hline

12$\times$Pix2Pix \cite{isola2017image}       	&171.3$\pm$0.4&	 {0.9206}$\pm$0.0013&	 {24.12}$\pm$0.02&	 {15.65}$\pm$0.16\\

4$\times$TCMGAN \cite{xin2020multi}	&168.6$\pm$0.2&	0.9413$\pm$0.0010&	24.87$\pm$0.01&	16.73$\pm$0.15\\
\hline\hline
Proposed-$\mathcal{L}_1^{disen}$            &{159.8}$\pm$0.3&	{0.9594}$\pm$0.0016& {25.21}$\pm$0.02&  {18.10}$\pm$0.13\\ 
Proposed              &\textbf{157.2}$\pm$0.2&	\textbf{0.9625}$\pm$0.0012& \textbf{25.92}$\pm$0.01&  \textbf{18.54}$\pm$0.15\\\hline
\end{tabular}}\vspace{-15pt}
\label{tabel:1} 
\end{table}

\vspace{-5pt}
\subsection{Quantitative Evaluations}\vspace{-5pt}
The synthesized images were expected to have realistic-looking textures, and to be structurally coherent with its corresponding ground truth images $x_y$.~For quantitative evaluation, we adopted the widely used metrics including mean L1 error, structural similarity index measure (SSIM), peak signal-to-noise ratio (PSNR), and inception score (IS). 

Table \ref{tabel:1} lists numerical comparisons between the proposed framework, Pix2Pix \cite{isola2017image}, and TCMGAM \cite{xin2020multi} for the 85 testing subjects using the BraTS'18 database. Of note, proposed-$\mathcal{L}_1^{disen}$ indicates the proposed model without the disentanglement constraint $\mathcal{L}_1^{disen}$. The proposed unified conditional disentanglement framework outperformed the other comparison methods w.r.t. these metrics, and the performance with the proposed framework was better than the proposed framework without the $\mathcal{L}_1^{disen}$. 

\begin{table}[t]
\centering
\caption{Comparisons of the segmentation accuracy} 
\resizebox{0.6\linewidth}{!}{
\begin{tabular}{c|c}
\hline
Methods& DICE \\\hline\hline
12$\times$Pix2Pix\cite{isola2017image}    &        0.7436$\pm$0.0017                	 \\ 
4$\times$TCMGAN\cite{xin2020multi}	& 0.7673$\pm$0.0014\\\hline\hline

Proposed-$\mathcal{L}_1^{disen}$ & 0.7791$\pm$0.0013\\ 
Proposed                                & \textbf{0.7862}$\pm$0.0015\\\hline\hline
Original 4 Modalities & 0.8142$\pm$0.0012  \\\hline
\end{tabular}
}
\label{tabel:2}\vspace{-15pt}
\end{table}

\subsection{Tumor Segmentation Results}\vspace{-5pt}
In Table \ref{tabel:2}, we followed \cite{xin2020multi} to use the synthesized images by different methods to boost the tumor segmentation accuracy. Specifically, we sampled a slice with any modality in the testing data, and used our unified translation framework to generate its complementary three modalities \cite{xin2020multi}. Then, we concatenated the real slice and its generated three modalities as input to the segmentor. We note that only using the additional generated complementary slices can achieve improvements over only using one real slice \cite{xin2020multi}. For example, the DICE score was 0.7404 using only T2-weighted MRI for segmentation. We also computed the DICE score using the entire four real modalities, which served as an ``upper bound".

The proposed unified conditional disentanglement framework yielded better segmentation performance than the baseline model Pix2Pix \cite{isola2017image} and TCMGAN \cite{xin2020multi}. In addition, the DICE score of our conditional disentanglement framework was close to the upper bound which was computed using four real modalities. It was seen from our experiments that using all of the pairs in  our training and the use of the cycle-constraint provided more accurate tumor shape recovery, thus leading to the better segmentation results.

\section{Conclusion}\vspace{-5pt}

This work presented a unified conditional disentanglement framework for co-registered multimodal translation based on a single set of target modality conditioned autoencoder and multi-task discriminator. The encoder is learned to extract the disentangled anatomical information by enforcing the consistency of two co-registered images with different modalities. The autoencoder is simply recalled twice to form a circular processing flow to enforce the cycle-constraint.~Our framework is scalable to many modalities and effective to utilize the entire paired training data.~In addition, our framework demonstrated superior performance on the tumor segmentation task over the compared methods using the generated images.

\section{COMPLIANCE WITH ETHICAL STANDARDS}\vspace{-10pt}
This research study was conducted retrospectively using human subject data made available in open access by
\href{https://www.med.upenn.edu/sbia/brats2018/data.html}{BraTS'18}. Ethical approval was not required as confirmed by the license attached with the open access data.

\section{ACKNOWLEDGMENTS}\vspace{-10pt}
{This work is partially supported by NIH R01DC018511, R01DE027989, and P41EB022544.}\vspace{-10pt}

\bibliographystyle{IEEEbib}
\bibliography{refs}

\end{document}